\documentclass[10pt]{article}
\usepackage{fullpage}
\usepackage{amsmath}
\usepackage{amssymb}
\usepackage[dvips]{epsfig}

\advance \textwidth by -2in
\advance \textheight by -3in
\def\##1{{\bf #1}}
\def\=#1{\underline{\underline #1}}
\def\~#1{{\tilde{\bf #1}}}

\def\eps{\epsilon}
\def\epso{\epsilon_0}
\def\muo{\mu_0}
\def\ko{k_0}
\def\co{c_0}

\def\.{\mbox{ \tiny{$^\bullet$} }}

\def\epsr{\epsilon_r}
\def\mur{\mu_r}

\def\curl{\nabla\times}

\def\les{\left[}
\def\ris{\right]}

\def\c#1{\cite{#1}}
\def\l#1{\label{#1}}
\def\r#1{(\ref{#1})}

\begin{document}

\vspace{5mm}

\noindent{\bf {\Large SIMPLE DERIVATION OF \\ DYADIC GREEN FUNCTIONS\\
OF A SIMPLY MOVING, ISOTROPIC,\\  DIELECTRIC--MAGNETIC MEDIUM }}

 \vspace{5mm} \large

\noindent
{\bf Akhlesh  Lakhtakia}$^1$\footnote{Corresponding author. E--mail: akhlesh@psu.edu}
and
{\bf Tom G. Mackay}$^2$\footnote{E--mail: T.Mackay@ed.ac.uk}\\

 \vspace{5mm}

\noindent {$^1$CATMAS~---~Computational \& Theoretical
Materials Sciences Group\\ Department of Engineering Science and
Mechanics\\ Pennsylvania State University\\ University Park, PA
16802--6812, USA}\\

 \vspace{5mm}

\noindent {$^2$School of Mathematics\\
University of Edinburgh\\ Edinburgh EH9 3JZ\\ United Kingdom}\\

\vspace{4mm}

\normalsize

\noindent {\bf ABSTRACT:}    Dyadic Green functions for
time--harmonic fields in a  homogeneous,
isotropic, dielectric--magnetic medium, moving with constant velocity, are derived by first
implementing a simple transformation and then using the 
dyadic Green functions
available for uniaxial dielectric--magnetic mediums.


\newpage

\normalsize

A significant portion of the electromagnetics literature is devoted
to  wave propagation in simply moving, homogeneous, isotropic,
dielectric--magnetic mediums \c{VBrel}. The dyadic Green functions (DGFs)
of such a medium (with respect to an observer in an inertial
non--co--moving frame of reference) have  therefore been derived,
and are available in closed form \c{Chen}. We present here an
alternative~---~and simple~---~derivation of these DGFs by first
implementing a straightforward transformation and then using the DGFs
available for uniaxial dielectric--magnetic mediums.

From the perspective of a co--moving observer, let  the frequency--domain
constitutive properties
of a medium be characterized by its 
relative permittivity $\eps_r$ and relative permeability $\mu_r$.
Suppose next that this medium  is moving with constant velocity
$\#v$ with respect to another inertial observer; then, the
 Minkowski constitutive relations of this medium
may be stated as follows \c{Chen}:
\begin{eqnarray}
\#D &=& \epso \epsr \, \=\alpha\.\#E +{c_0^{-1}}\#m
\times\#H \,,
\\[6pt]
\#B &=&  -\, {c_0^{-1}} {\#m\times\#E} + \muo
\mur \, \=\alpha\.\#H
\,.
\end{eqnarray}
Here and hereafter,  
\begin{eqnarray}
\#v&=&v\hat{\#v}\,,\\[5pt]
\=\alpha &=& \alpha\,\=I + (1-\alpha)\,\hat{\#v}\hat{\#v} \,,\\[5pt]
\alpha & = & \frac{1-\beta^2}{1- \epsr \mur \beta^2} \,,\\[5pt]
\#m&=&m\hat{\#v} \,,\\[5pt]
m &=& \beta\,\frac{ \epsr \mur -1} {1- \epsr \mur \beta^2}\,,\\[5pt]
\beta&=&v/\co\,,
\end{eqnarray}
and $\co=1/\sqrt{\epso\muo}$ is the speed of light in free space (i.e., vacuum).

The frequency--domain Maxwell curl postulates in the chosen medium
(with respect to the non--co--moving observer)
 may be
set down
as
\begin{eqnarray}
\label{Eeqn}
&&\curl\#E(\#r)=i\omega\les \muo\mur\,\=\alpha\.\#H(\#r) - {c_0^{-1}}\,\#m\times\#E(\#r)\ris \,,\\[5pt]
&&\curl\#H(\#r)=-i\omega\les
\epso\epsr\,\=\alpha\.\#E(\#r) + {c_0^{-1}}\,\#m\times\#H(\#r)\ris
+\,\#J(\#r)\,,
\label{Heqn}
\end{eqnarray}
where the phasor $\#J(\#r)$ represents the source electric current density.
Our objective is to find the DGFs
$\=G_{\,e}(\#r,\#s)$ and $\=G_{\,m}(\#r,\#s)$ such that
\begin{eqnarray}
\label{Esol} \#E(\#r) &=&
i\omega\muo\int\int\int\,\=G_{\,e}(\#r,\#s)\.\#J(\#s)\,d^3\#s\,,
\\[5pt]
\label{Hsol}
 \#H(\#r)
&=&\int\int\int\,\=G_{\,m}(\#r,\#s)\.\#J(\#s)\,d^3\#s\,,
\end{eqnarray}
with  the integrations being  carried out over the region where 
 $\#J(\#r)$ is nonzero.

We begin by using a transformation provided by Lakhtakia \& Weiglhofer \c{LW97};
thus, we define 
 transformed field and source  phasors as
\begin{eqnarray}
\label{tr1}
&&\#e(\#r)=\#E(\#r)\,\exp(ik_0\,\#m\.\#r)\,, \l{e_LW}
\\[5pt]
\label{tr2}
&&\#h(\#r)=\#H(\#r)\,\exp(ik_0\,\#m\.\#r)\,,
\\[5pt]
\label{tr3}
&&\#j(\#r)=\#J(\#r)\,\exp(ik_0\,\#m\.\#r)\,, \l{j_LW}
\end{eqnarray}
where $\ko=\omega\sqrt{\epso\muo}$ is the free--space wavenumber.
Hence, \r{Eeqn} and \r{Heqn} respectively transform to
\begin{eqnarray}
\label{eeqn}
&&\curl\#e(\#r)=i\omega  \muo\mur\,\=\alpha\.\#h(\#r)  \,,
\\[5pt]
\label{heqn}
&&\curl\#h(\#r)=-i\omega
\epso\epsr\,\=\alpha\.\#e(\#r)
+\#j(\#r)\,.
\end{eqnarray}

The solutions of \r{eeqn} and \r{heqn} may be written as
\begin{eqnarray}
\#e(\#r)&=&i\omega\muo\int\int\int \=g_e(\#r,\#s)\.\#j(\#s)\,d^3\#s\,,
\\
\#h(\#r)&=&\int\int\int \=g_m(\#r,\#s)\.\#j(\#s)\,d^3\#s\,,
\end{eqnarray}
The DGFs appearing in these solutions are for a dielectric--magnetic
medium with a distinguished axis, and are available as \c{LVV}
\begin{eqnarray}
\=g_e(\#r,\#s)&=&\mur\alpha \left(\=\alpha^{-1} +\frac{\nabla\nabla}{k_v^2}
\right)\,\frac{\exp(ik_v R_v)}{4\pi R_v}\,,
\\[5pt]
\=g_m(\#r,\#s)&=&\epsr\alpha\,\=\alpha^{-1}\times\nabla
\frac{\exp(ik_v R_v)}{4\pi R_v}\,,
\end{eqnarray}
where
\begin{eqnarray}
k_v&=&\ko\alpha\sqrt{\epsr\mur}
\,,\\[5pt]
R_v&=&\sqrt{(\#r-\#s)\.\=\alpha^{-1}\.(\#r-\#s)}\,.
\end{eqnarray}

Reversing the transformation \r{tr1}--\r{tr3}, we find the sought--after DGFs
as follows:
\begin{eqnarray}
\nonumber
\=G_e(\#r,\#s)&=&\mur\alpha\,\exp\left[-i\ko\#m\.(\#r-\#s)\right]\,
\\[5pt]
&&\,
 \left(\=\alpha^{-1} +\frac{\nabla\nabla}{k_v^2}
\right)\,\frac{\exp(ik_v R_v)}{4\pi R_v}\,,
\\[6pt]
\nonumber
\=G_m(\#r,\#s)&=&\epsr\alpha\,\exp\left[-i\ko\#m\.(\#r-\#s)\right]\,\\[5pt]
&&\,
\=\alpha^{-1}\times\nabla
\frac{\exp(ik_v R_v)}{4\pi R_v}\,.
\end{eqnarray}
These expressions can be put in the same form as the DGFs
provided in Chen's textbook \cite[Chap 11]{Chen}.


\begin{thebibliography}{99}

\bibitem{VBrel}
J. Van Bladel, Relativity and engineering, Springer, Berlin, 1984.


\bibitem{Chen}
H.C. Chen, Theory of electromagnetic waves, McGraw--Hill, New York, 1983.

\bibitem{LW97}
A. Lakhtakia and W.S. Weiglhofer,
On electromagnetic fields in a linear medium with gyrotropic--like
magnetoelectric properties,
{Microwave Opt Technol Lett\/} {15} (1997) 168--170.

\bibitem{LVV}
A. Lakhtakia, V.V. Varadan and V.K. Varadan,
Time--harmonic and time--dependent dyadic Green's functions for
some uniaxial gyro\-electro\-magnetic media,
Appl Opt 28 (1989) 1049--1052.



\end{thebibliography}
\end{document}